\documentclass[a4paper, amsfonts, amssymb, amsmath, reprint, nofootinbib,superscriptaddress,twoside]{revtex4-2}

\usepackage{algcompatible}
\usepackage{newfloat}

\DeclareFloatingEnvironment[
    fileext=loa,
    listname=List of Algorithms,
    name=ALGORITHM,
    placement=tbhp,
]{algorithm}
\usepackage{algpseudocode}
\usepackage{lipsum}
\usepackage{braket}
\usepackage{comment}

\usepackage{graphicx}
\usepackage{dcolumn}
\usepackage{xcolor}
\usepackage{lineno}

\usepackage{bm}

\usepackage{hyperref}
\hypersetup{
    colorlinks=true,
    linkcolor=blue,
    filecolor=magenta,      
    urlcolor=cyan,
}

\begin{document}

\title{Quantum data generation in a denoising model with multiscale entanglement renormalization network}

\author{Wei-Wei Zhang}\footnote{corresponding authors:}
\email{wei-wei.zhang@nwpu.edu.cn}
\affiliation{%
School of Computer Science, Northwestern Polytechnical University, Xi’an 710129, China 
}%
\author{Xiaopeng Huang}%
\affiliation{%
school of cybersecurity, Northwestern Polytechnical University, Xi’an 710129, China 
}%
\author{Shenglin Shan}
\affiliation{%
School of Artificial Intelligence, Optics and Electronics (iOPEN), Northwestern Polytechnical University, Xi’an 710072, China 
}%
\author{Wei Zhao}
\email{zhaowei9801@163.com}
\affiliation{
National Key Laboratory of Security Communication,  Chengdu 610041, China 
}
\author{Beiya Yang}%
\affiliation{%
School of Computer Science, Northwestern Polytechnical University, Xi’an 710129, China 
}%
\author{Wei Pan}%
\affiliation{%
School of Computer Science, Northwestern Polytechnical University, Xi’an 710129, China 
}%
\author{Haobin Shi}%
\affiliation{%
School of Computer Science, Northwestern Polytechnical University, Xi’an 710129, China 
}%

\date{\today}

\begin{abstract}
Quantum technology has entered the era of noisy intermediate-scale quantum (NISQ) information processing.
The technological revolution of machine learning represented by generative models heralds a great prospect of artificial intelligence, and the huge amount of data processes poses a big challenge to existing computers. The generation of large quantities of quantum data will be a challenge for quantum artificial intelligence. 
In this work, we present an efficient noise-resistant quantum data generation method that can be applied to various types of NISQ quantum processors, where the target quantum data belongs to a certain class and our proposal enables the generation of various quantum data belonging to the target class. Specifically, we propose a quantum denoising  probability model (QDM) based on a multiscale entanglement renormalization network (MERA) for the generation of quantum data. To show the feasibility and practicality of our scheme, we demonstrate the generations of the classes of  GHZ-like states and  W-like states with a success rate above 99\%. Our MREA QDM can also be used to denoise multiple types of quantum data simultaneously.  We show the success rate of denoising both GHZ-like and W-like states with single qubit noise environment of noise level within $1/4$  can approximate to be $100\%$, and with two other types of noise environment with noise level within $1/4$ can be above $90\%$. 
Our quantum data generation scheme provides new ideas and prospects for quantum generative models in the NISQ era.

\end{abstract}

\maketitle


\section{Introduction}
The development of quantum technology has entered a noisy intermediate-scale quantum (NISQ) era~\cite{preskill2018quantum}, and NISQ quantum processors have demonstrated their advantages in various areas~\cite{arute2019quantum,zhong2020quantum,baldwin2022re,zhu2022quantum,cao2023generation}. While quantum technology shows its computational advantage in complex problems for classical computers, manufacturing quantum computers remains challenging because of the unavoidable system noise and the  limited  amount of available qubits and types of operations on the chips~\cite{corcoles2019challenges,yi2024robust}.

Quantum state preparation is an important part of quantum computation and information processing and is still a challenging task~\cite{kolesnikow2024gottesman,catstates-ww,Yuan2023optimalcontrolled}. 
The quantum gates that are physically realizable by controlling the qubits in quantum hardware are defined as quantum instruction sets (QIS). Compiling quantum circuits into the product of the gates in a properly defined QIS is a fundamental step in quantum computing and 
quantum   compiling~\cite{hormozi2007topological,almudever2017engineering,venturelli2019quantum,sharma2020noise,gleinig2021efficient,zhang2022quantum,yuan2023optimal,melnikov2023quantum,lu2023quantum}. Because of the restrictions of the quantum operations realized on physical quantum chips, the general quantum state preparation strategy is still a challenging problem. The exploration of quantum state preparation using machine learning techniques has been proposed and proved to be promising~\cite{porotti2022deep, PhysRevA.108.052418,mackeprang2020reinforcement,zhang2019does,arrazola2019machine}.

Emergent Abilities demonstrated by Large Language Models such as ChatGPT have unleashed an AI revolution~\cite{10137850}.
Generative modeling is an unsupervised form of machine learning where the model learns to discover the patterns in input data. Using the learned knowledge, the model can generate new data on its own, which is relatable to the original training dataset~\cite{cao2023survey}. The main goal of generative models is to effectively and automatically generate new data satisfying certain conditions. 
Models like variational autoencoders (VAEs)~\cite{rezende2014stochastic}, generative adversarial network  (GANs)\cite{creswell2018generative}, and flow-based models~\cite{rezende2015variational} proved to be a great success in generating high-quality content, especially images. Diffusion models are a new type of generative model that has proven to be better than previous approaches~\cite{ho2020denoising}. Denoising diffusion probabilistic models are currently becoming the leading
paradigm of generative modeling for many important data modalities~\cite{ho2020denoising}. 
With the boost of generative AI, its requirement for computation power has become a bottleneck, and the quantum computation power attracts more attention in this regard.

\begin{figure}[ht]
\includegraphics[width=\columnwidth]{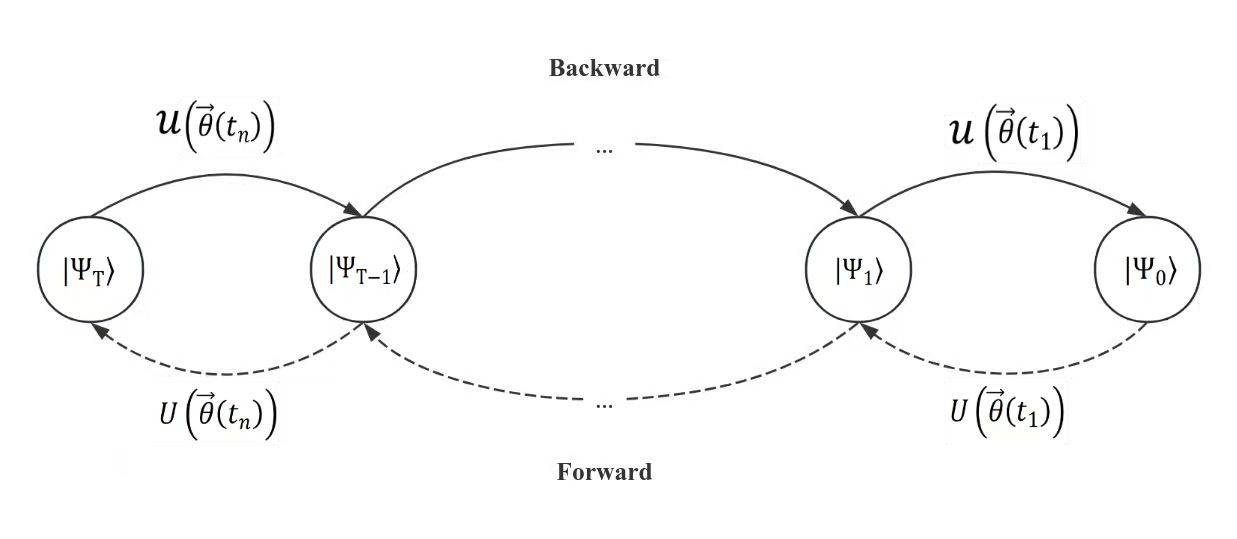}
\caption{Sketch of our quantum data generation scheme.} 
\label{fig:sketch-1type}
\end{figure}

With the rapid development of quantum technology and its applications in various scenarios, the requirement for quantum arbitrary state preparations is getting urgent. 
In this work, we propose an efficient noise-resistant method for quantum data generation based on the quantum denoising process. Our method is compatible with all types of NISQ hardware devices for arbitrary quantum state preparation. Specifically, we
propose a Quantum Denoising probability Model (QDM) based on a Multiscale Entanglement Renormalization Network (MERA) for the generation of large quantities of new quantum data belonging to the same class of training data. To show the feasibility and practicality of our method, 
we demonstrate the generations of GHZ-like states and W-like states with the corresponding average success rate reaching  99.20\% and 99.05\%  respectively. 
Our MREA QDM can also be used to denoise multiple types of quantum data
simultaneously. We show the success rate of denoising both GHZ-like and W-like states with the single qubit Rx/Ry
noise level of $\frac{\pi}{4}$  using the same network reaching approximately 100\%. For the two extra complex noise environments with noise level within $\pi/4$, our scheme can denoise and recover the original data with fidelity above 90\%, which shows the generality of our scheme.
Our quantum data generation scheme provides new ideas and prospects for quantum computation and quantum 
 information processing.

\section{Quantum data generation with  denoising model}
\subsection{The model}
In this section, we propose our quantum data generation scheme, which is a quantum  analogy of classical denoising
diffusion probability model (DDPM)~\cite{ho2020denoising}. Inspired by non-equilibrium thermodynamics, the classical generative diffusion models define a Markov chain of diffusion steps to slowly add random noise to data and then learn to reverse the diffusion process to construct desired data samples from the noise, which are named as forward and backward processes. 
The spirit of classical DDPM is nonequilibrium thermodynamics, and it learns to generate by denoising. During the forward process, scheduled noise is added to the data until it becomes full noise data. During the backward process, the generated full noise data is trained to converge to the original data class which is a denoising process. 

Inspired by DDPM, our quantum generative scheme is a composite of forward and backward processes as shown in the sketch of our scheme in Fig.~\ref{fig:sketch-1type}. Our forward process is realized with a scheduled random circuit which incorporates an element of randomness into the local unitary operations and measurements of a quantum circuit.  The idea is similar to that of random matrix theory which is to use the quantum random circuit to obtain almost exact results of non-integrable, hard-to-solve problems by averaging over an ensemble of outcomes. The incorporation of randomness into the circuits has been utilized in the validation of quantum computers, which is the method that Google used when they claimed quantum supremacy in 2019~\cite{arute2019quantum} and understanding the universal structure of non-equilibrium and thermalization processes in quantum many-body dynamics~\cite{nahum2017quantum,zhou2020entanglement}. 
Our backward process is a classical-quantum hybrid machine-learning process utilizing the gradient descent method to denoise the noise data into pure data.

During the forward process, the evolutionary quantum state will gradually lose its information, and the probability that the data belonging to the target type gradually drops to almost 0. During the backward process, the full noise data will gradually converge to the target type, and the probability that the data belonging to the target type increases to almost 1. 
 The forward process maps quantum data into noisy data out of the target data type space, and the backward process maps random noise data back into the target data type space.

In the forward process, random noise is scheduled to be slowly added to the quantum data during the evolution steps  $(i=1, 2, ..., n )$, and at the end of the forward process the samples are turned into approxited pure quantum noise data which is a quantum state with all the components equally weighted. The forward process can be written as

\begin{align}
    \ket{\Psi(n)}=U( \overrightarrow{\theta}( n))U(\overrightarrow{\theta}({n-1}))\dots U(\overrightarrow{\theta}(1))\ket{\Psi(0)}
\end{align}

where the formula of U is a composite of Rx, Ry, and Rz gates, \{$\overrightarrow{\theta}(i), i\in[1,n]$\} is the scheduled parameters for the random circuits during the forward process, $\Psi(0)$  and $\Psi(n)$ is the quantum data and the obtained full noise quantum data. In our simulations, we set the schedule for the parameters $\overrightarrow\theta(i)$ sampling from the scheduled Gaussian distributions. 

\begin{align}
\theta(i)=\frac{i}{n}\frac{\pi}{3}x;
i\in[1,n];x\sim\mathcal{N}(0,1)
\label{eq:gaussian-theta}
\end{align}

with the $\mathcal{N}(0,1)$ as the standard Gaussian distribution, the variable follows $\frac{\pi}{3}x\in[-\pi,\pi]$ with probability of 99.7\%.  Therefore, $\theta(i)$ defined in Eq.~\ref{eq:gaussian-theta} follows $\theta(i)\in[-\frac{i}{n}\pi,\frac{i}{n}\pi]$ with probability of 99.7\%. In our simulation of 8 qubit states generations, we set $n=16$; in our simulation of 16 qubit states  generation, we set $n=32$.

In the backward process, we train a quantum network to eliminate the added noise gradually. We start with full noise data (the last step of the forward process) and try to denoise the samples in the backward direction $(i=n, i={n-1}, ..., i=1)$ by adjusting our quantum network parameters $\overrightarrow\theta$. The backward process can be written as 

\begin{align}
\ket{\tilde{\Psi}(n)}=\mathcal{U}( \overrightarrow{\theta}( 1))\mathcal{U}(\overrightarrow{\theta}({2}))\dots \mathcal{U}(\overrightarrow{\theta}(n))\ket{\tilde{\Psi}(0)}
\end{align}
where \{$\overrightarrow{\theta}(i), i\in[1,n]$\} is the network parameters, $\tilde{\Psi}(0)$  and $\tilde{\Psi}(n)$ is the initial random full noise quantum data and corresponding denoised quantum data in one training round separately. The structure of the network $\mathcal{U}$ and the training loss function will be introduced in detail later. 
The pseudocode for network training in the backward process is shown in Alg.~\ref{alg:train}. 

By noting the trained network parameters as $\overrightarrow{\Theta}$, the target quantum data-generating process can be presented as

\begin{align}
\ket{\tilde{\Phi}(n)}=\mathcal{U}( \overrightarrow{\Theta}( 1))\mathcal{U}(\overrightarrow{\Theta}({2}))\dots \mathcal{U}(\overrightarrow{\Theta}(n))\ket{\tilde{\Phi}(0)}
\end{align}

with $\ket{\tilde{\Phi}(0)}$ as the initial full noise quantum data, the $\Theta(i), i\in\{1,\dots,n\}$ follows a scheduled formula. The pseudocode for target quantum data generation  is shown in Alg.~\ref{alg:gene}.

\begin{algorithm}
\caption{Training}\label{alg:train}
\begin{algorithmic}[1]

\State \textbf{Repeat:}
\State $\Psi(n) \sim$ \text{full noise states generated with target states} 

\State $\theta(i) \sim$ Eq.(\ref{eq:gaussian-theta})
\State $|\Psi(n-1)\rangle=U(\overrightarrow{\theta}(n)) |\Psi(n)\rangle $
\State Take gradient step on:
$\nabla_\theta\left\|\boldsymbol{|\Psi(n)\rangle}-|\boldsymbol{\Psi(n-1)\rangle} \right\|^2$
\State $\textbf{until}$ converged
\end{algorithmic}
\end{algorithm}

\begin{algorithm}
\caption{Data generating}\label{alg:gene}
\begin{algorithmic}[1]
\State \textbf{START:}
\State $\ket{\tilde{\Phi}(0)}\sim$ \text{full noise state}
\State for T,...,1 do
\State \quad$\overrightarrow{\Theta} \sim$ \text{Scheduled formula obtained from Alg.~\ref{alg:train}}
\State \quad$|\tilde{\Phi}(n)\rangle=\mathcal{U}(\overrightarrow{\Theta}(n)) 
|\tilde{\Phi}(n-1)\rangle $
\State end for
\end{algorithmic}
\end{algorithm}

\begin{figure}[ht]
\includegraphics[width=\columnwidth]{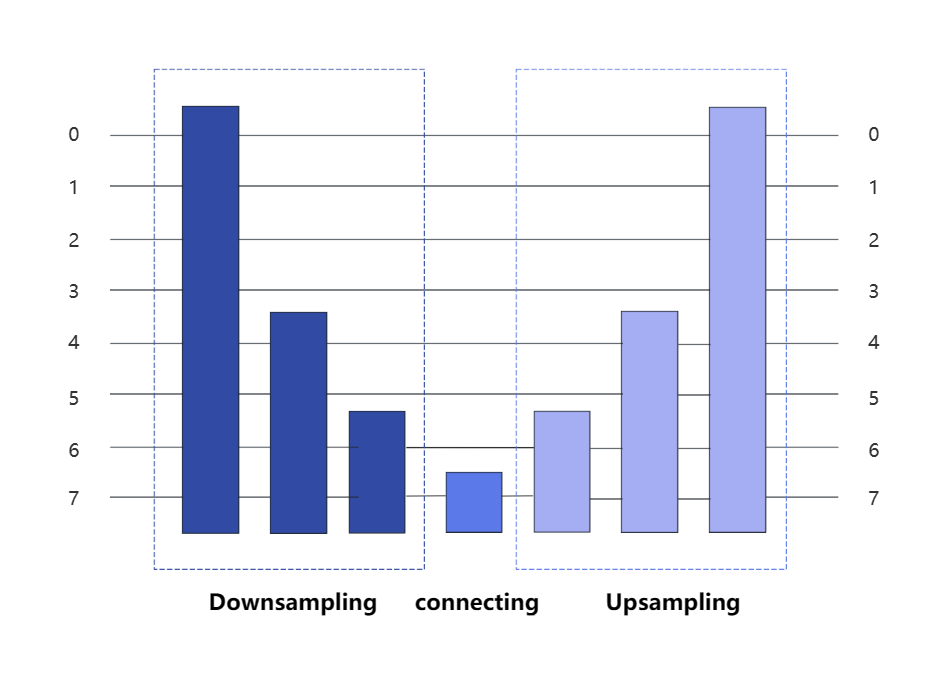}
\caption{A sketch of QUNET structure for the quantum data generation. In the left part of QUNET, the dark blue blocks consist of a quantum convolution block, a quantum noise block, and a quantum pooling block; in the right part of QUNET, the light blue blocks consist of a quantum convolution block and a quantum noise block.} 
\label{qunet-circuit}
\end{figure}

The network we design is the quantum analog of the classical UNET structure. UNET is a U-shaped encoder-decoder network architecture, which was designed for Biomedical Image Segmentation and originally consists of four encoder blocks as downsampling path and four decoder blocks as upsampling path connected via a bridge~\cite{ronneberger2015u}. As the demonstration of good performance in quantum error correction, we design our denoising network as the U shape, named QUNET. 

The design of our QUNET  is inspired by the multiscale entanglement
renormalization network (MERA), which belongs to a class of variational ansatz for quantum many-body states and has been used for quantum convolution neural network~\cite{cong2019quantum}. 
MERA is based on a real-space renormalization group procedure called entanglement renormalization,
designed to systematically handle entanglement at different length scales along the coarse-graining
flow~\cite{vidal2008class}. The high-dimensional and complex quantum state is "coarse-grained", and the properties such as entanglement and causality in the system are equivalently condensed into the reduced-dimensional state, and the system properties
for analysis.

\begin{figure}[t]
\includegraphics[width=\columnwidth]{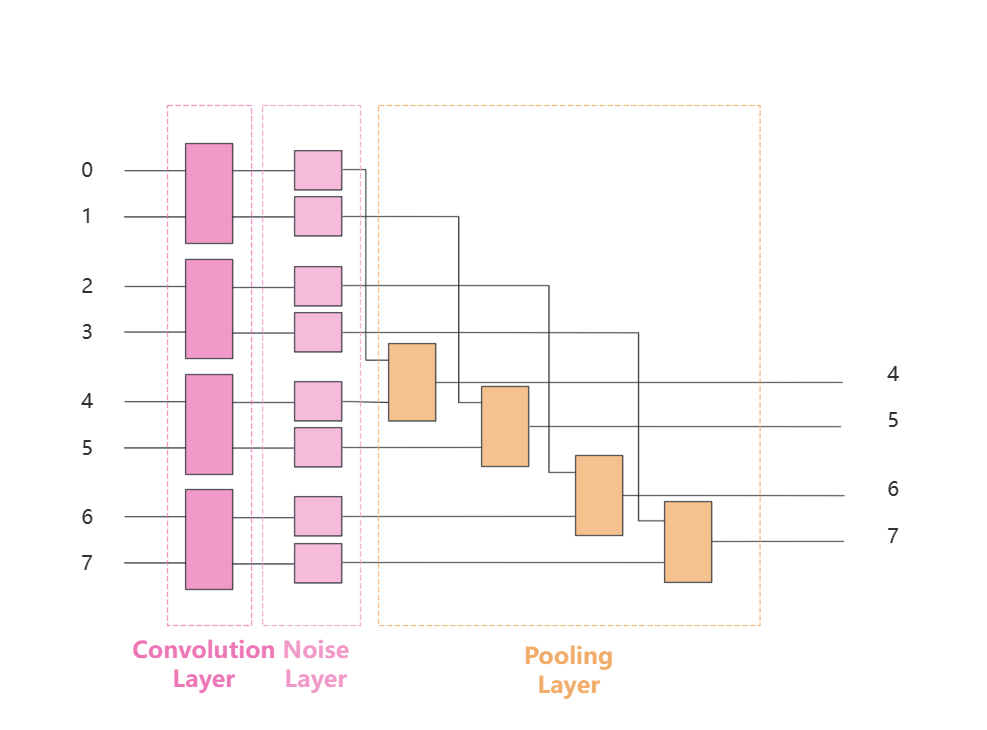}
\caption{The detailed building blocks of an 8 qubit QUNET's downsampling path in Fig~\ref{qunet-circuit}.} 
\label{qunet-circuit-detail}
\end{figure}
Our QUNET is a fully convolutional neural network composed of MERA,  that is designed to learn the target data properties from few training samples. we take the 8-qubit states for the demonstrations except in the cases where the specific size is specified. The sketch of our QUNET is shown in Fig.~\ref{qunet-circuit}, the encoder network is the left part of the circuit, i.e. the downsampling path, which has half the spatial dimensions and double the number of filters (feature channels) at each encoder block. Likewise, the right part of the circuit is 
 the decoder network, i.e. the upsampling path, which doubles the spatial dimensions and half the number of feature channels. The details of the UNET blocks are shown in Fig.~\ref{qunet-circuit-detail}, which is the quantum circuit for the downsampling part of QUNET, and the upsampling part is the structure of its mirror symmetry. 

 Specifically,  the coarse-graining flow of MERA plays the role of pooling layers by extracting the feature map of QUNET. The MERA representation of QUNET is a bridged structure, where the left part of the bridge is the downsampling path and the right part of the bridge is upsampling path. Our QUNET MERA structure is constructed with two types of operations named disentanglers (the pink blocks) and isometries (the orange blocks). The disentanglers play the role of convolution layers and the isometries play the role of pooling layers.

The training process of our method is a classical-quantum hybrid process. During the training process, we measure the outcomes of QUNET, and obtain the probability distributions of each measurement basis. The loss function is set as the mean absolute error between the measured probabilities and the target probabilities. We use the classical Adam algorithm to optimize the quantum circuit parameters.

\subsection{Our experimental  demonstration for the generation of quantum data}

Our scheme is a general scheme for quantum data preparation which can be used for the generation of arbitrary types of quantum states. In this section, we demonstrate the generation of two types of quantum data to show the functionality of our scheme, generalized GHZ states named GHZ-like states and the generalized W states named W-like states.

\begin{figure}[t]
\includegraphics[width=\columnwidth]{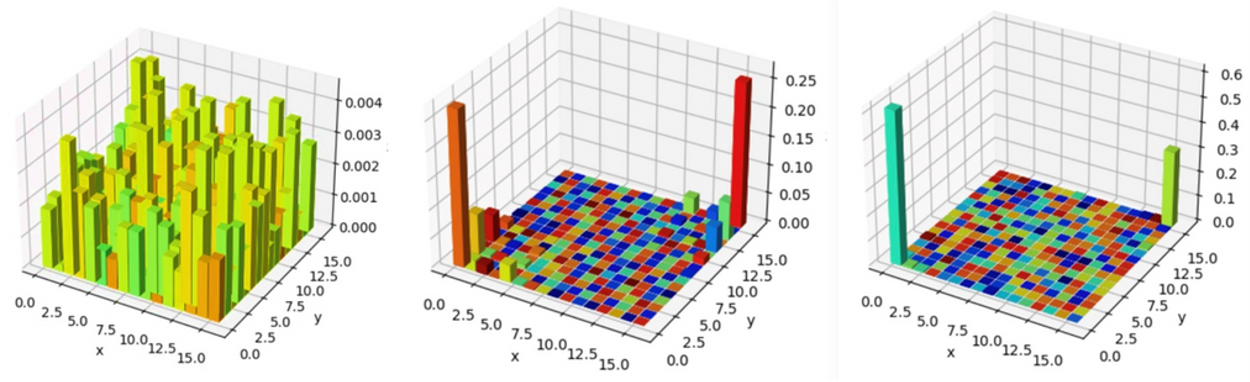}\\
(a)\\
\includegraphics[width=\columnwidth]{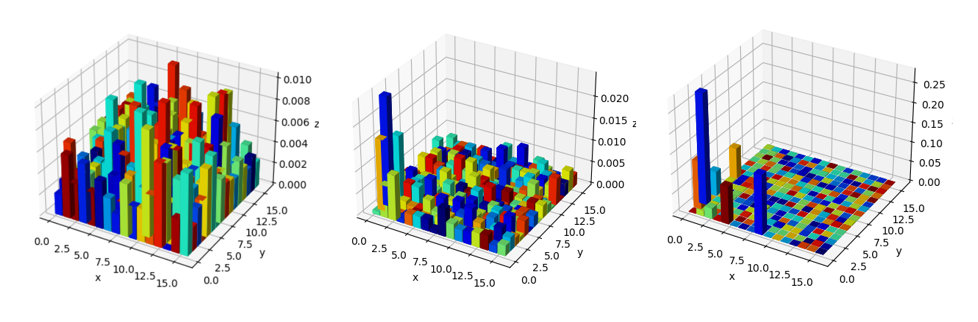}\\
(b)
\caption{The generation process of an 8 qubit GHZ-like (a) state and W-like state (b) with the QUNET in Fig.~\ref{qunet-circuit}.} 
\label{generate-process}
\end{figure}

The target types of quantum states to be generated are the GHZ-like states and W-like states, which are the quantum state types holding a similar measurement probability distribution to the standard GHZ state and W state. Specifically, the measurement probability distribution of GHZ-like states has only the nonzero probabilities at the measurement basis $\ket{00000000}$ and $\ket{00000000}$. Likewise, the measurement probability distribution of W-like states has only the nonzero probabilities at the measurement basis {$\ket{00000001}$, $\ket{00000010}$, $\ket{00000100}$, $\ket{00001000}$, $\ket{00010000}$, $\ket{00100000}$, $\ket{01000000}$, $\ket{10000000}$ }. 
The mathematical definition of GHZ-like states $\ket{\widetilde{\text{GHZ}}}$ and W-like states $\ket{\widetilde{\text{W}}}$ is following  
\begin{equation}  \widetilde{\text{GHZ}}=\alpha|00000000\rangle+\beta|11111111\rangle
\label{eq:ghz-1}
\end{equation}
with $\alpha$,$\beta$,
 as complex numbers and  $|\alpha|^2+|\beta|^2=1$. 
\begin{equation}
\begin{split}
\widetilde{\text{W}}=&\alpha_1|10000000\rangle+\alpha_2|01000000\rangle\\+&\alpha_3|00100000\rangle+\alpha_4|00010000\rangle\\+&\alpha_5|00001000\rangle+\alpha_6|00000100\rangle\\+&\alpha_7|00000010\rangle+\alpha_8|00000001\rangle
\end{split}  
\label{eq:w-1}
\end{equation}  
with 
$\alpha_{i}, i\in[1,...,8]$ as complex numbers and  $\sum_i|\alpha_i|^2=1$. 

To visualize a quantum data generation process, we map a $n$ qubit quantum state into a classical 2D-array size of $2^{\lceil n/2\rceil}\times2^{\lceil n/2\rceil}$ and present it as a 2D image.  The coordinates of classical images encode the quantum computational basis. In the 8-qubit case, a quantum state is encoded into a 2D image with the size of $16\times 16$ as shown in Fig.~\ref{generate-process}. The column height of the histogram encodes the probability value that the quantum state is in the corresponding quantum basis. 
Specifically,  the coordinate (1,1) in our representation corresponds to the computational basis $|00000000\rangle$, and The coordinate (16,16) corresponds to the computational basis $|11111111\rangle$. The generation process of a GHZ-like state and a W-like state with a trained QUNET  is shown in Fig.~\ref{generate-process}, where the initial full noise quantum states (the first column in Fig.~\ref{generate-process}) are transformed into a GHZ-like state or a W-like state (the third column in Fig.~\ref{generate-process}) by the denoising process of our QDM. The middle column images in Fig.~\ref{generate-process} show the quantum state during the process, where the main components of GHZ-like state and W-like state are kept and the other components are gradually suppressed during the generation process. 

Our quantum data generation process is a quantum denoising process where the network is trained to learn the mapping rule from a full-noise quantum data space to the target quantum data space. The size of training data for our 8(16)-qubit states generation is 500 (1000). Different arbitrary noise quantum data will be converted into different quantum data belonging to the same type, which offers an efficient quantum data generation scheme.

\begin{figure}[t]
(a)\includegraphics[width=0.9\columnwidth]{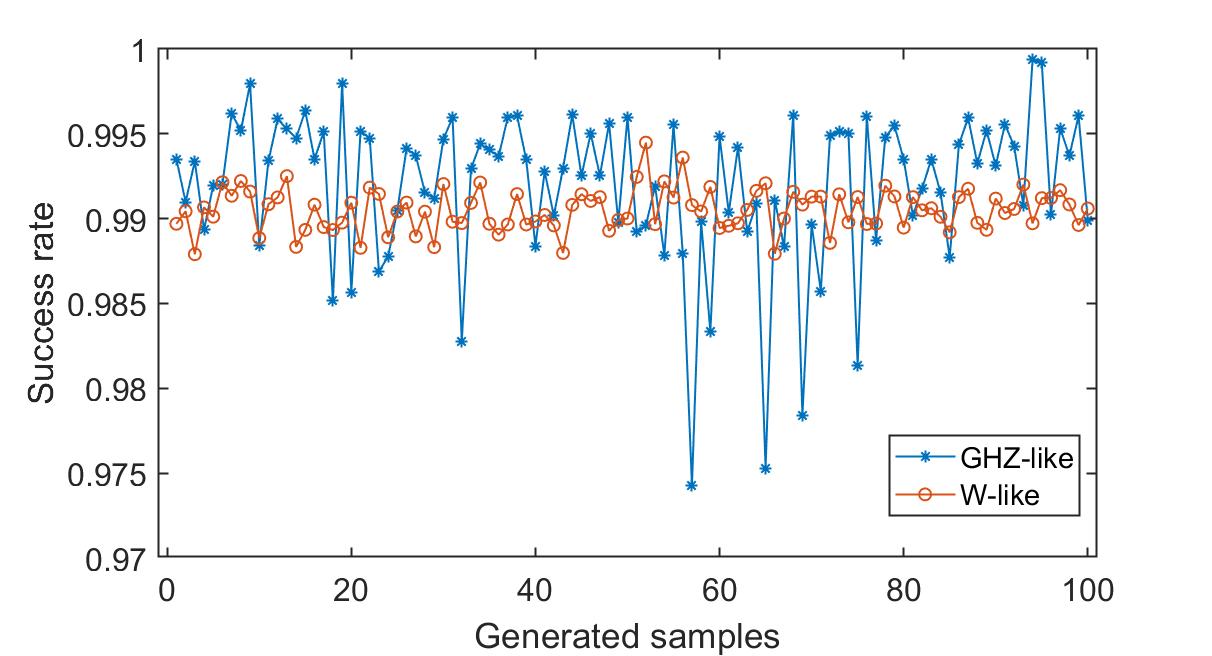}
(b)\includegraphics[width=0.88\columnwidth]{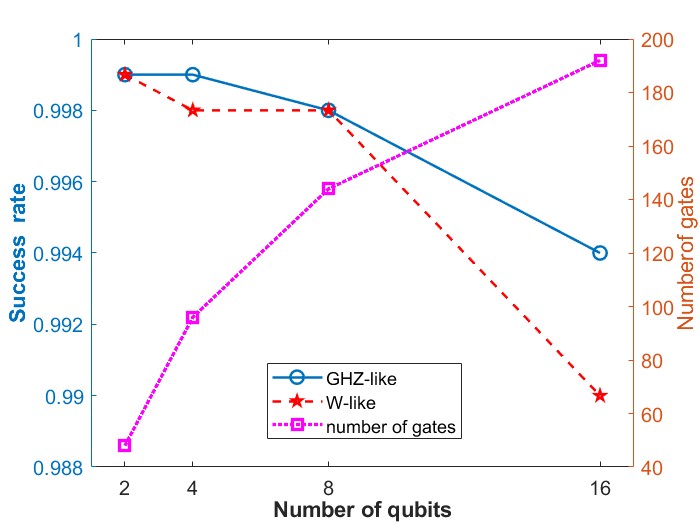}
\caption{The generation success rate of generating a hundred samples of  (a) 8 qubit  GHZ-like and W-like states with our scheme (b) The relation between the data generation success rate and the size of the generated state (dot for GHZ-like states and star for W-like states), and the relation between the number of the gates in our network and the size of the generated state. } 
\label{success-rate}
\end{figure}

The success rate of our quantum data generation scheme is evaluated with the probability that the generated quantum data belongs to the target data type. In our demonstration,  the probability that the generated data belongs to GHZ-like states or W-like states is defined as follows, where we name it as the success rate.

\begin{equation}
P_\textbf{succ}^{\text{GHZ}}=P_\text{00000000}+P_\textbf{111111111}.
\end{equation}
\begin{equation}
\begin{split}
P_\textbf{succ}^{\text{W}}=&P_\text{10000000}+P_\text{01000000}\\+&P_\text{00100000}+P_\text{00010000}\\+&P_\text{00001000}+P_\text{00000100}\\+&P_\text{00000010}+P_\text{00000001}
\end{split}
\end{equation}
Here $P_\text{00000000}$, $P_\text{111111111}$, $P_\text{10000000}$, $P_\text{01000000}$, $P_\text{00100000}$, $P_\text{00010000}$, $P_\text{00001000}$, $P_\text{00000100}$, $P_\text{00000010}$ and  $P_\text{00000001}$ are the probabilities of the generated states project to the corresponding computation basis \{$\ket{00000000}$, $\ket{11111111}$, $\ket{00000001}$, $\ket{00000010}$, 
$\ket{00000100}$, 
 $\ket{00001000}$, $\ket{00010000}$, $\ket{00100000}$, $\ket{01000000}$, $\ket{10000000}$\}. 
 
Our simulations  in 
Fig.~\ref{success-rate} shows the values of $P_\text{succ}$ of our generated 100 8 qubit and 16 qubit GHZ-like and W-like states. 
We use the average success rate of the generated quantum data to quantify the quality of our generation network. 
\begin{align}
    Q_\text{GHZ}=&\frac{1}{N}\sum_{i=1}^NP_\text{succ}^\text{GHZ}(i)\\
     Q_\text{W}=&\frac{1}{N}\sum_{i=1}^NP_\text{succ}^\text{W}(i)
\end{align}
where, $P_\text{succ}^\text{GHZ}(i)$ and $P_\text{succ}^\text{W}(i)$ are the success probability of the $i\text{th}$ generated GHZ-like state and W-like state, $N$ the the total number of the generated quantum data. The closer the value is to 1, the better the generation effect. In our experimental simulations in~Fig.~\ref{success-rate}, the average success rate reaches 99.2\% and 99.05\% for the generated 8 qubit GHZ-like states and  W-like states, 99.4\% and 99.05\% for the generated 16 qubit GHZ-like states and  W-like states. Fig.~\ref{success-rate} shows the scalability of our model by showing the relation between the data generation success rate and the size of the generated state (dot for GHZ-like states and star for W-like states), and the relation between the number of the gates in our network and the size of the generated state. We can see that with the increase of the number of qubits in the generated states, the success rate does not drop too much and the number of the gates required in our network is increased following $\text{O}(\log N)$ (note: the horizontal index is increased exponentially).

\begin{table*}[]
    \centering
    \begin{tabular}{c|c|c|c|c|c|c}\hline\hline
       Data type &  SNR(B) & SNR(A) & MSE (B)  &MSE (A) &PSNR (B)  & PSNR(A) \\\hline
       GHZ-Rx  &-1.31& 13.96 &0.48 &0.0051 &51.33 &71.14  \\\hline
       W-Rx  & 4.91 &12.12 &0.49& 0.0241 &51.25 &64.35 \\\hline
       GHZ-Ry & -0.09 &12.97 &0.59 &0.0488 &50.43 &61.28 \\\hline
       W-Ry  &4.89 &11.59 &00.73& 0.0662 &49.56 &59.96 \\\hline
             GHZ-$\epsilon_\text{III}$ & -0.35 &11.6&0.79&0.07 &49.20&59.98\\\hline       W-$\epsilon_\text{III}$  & 4.22 &8.17 &0.98
       &0.15 &48.25 &56.4\\\hline\hline
    \end{tabular}
    \caption{The comparison of our denoising results, where `B' stands for before the denoising process and `A' stands for after the denoising process. SNR represents signal-noise ratio, MSE represents mean squared error, PSNR represents peak signal-noise ratio.}\label{table:ExpCom}
\end{table*}


\begin{figure}[ht]
\includegraphics[width=\columnwidth]{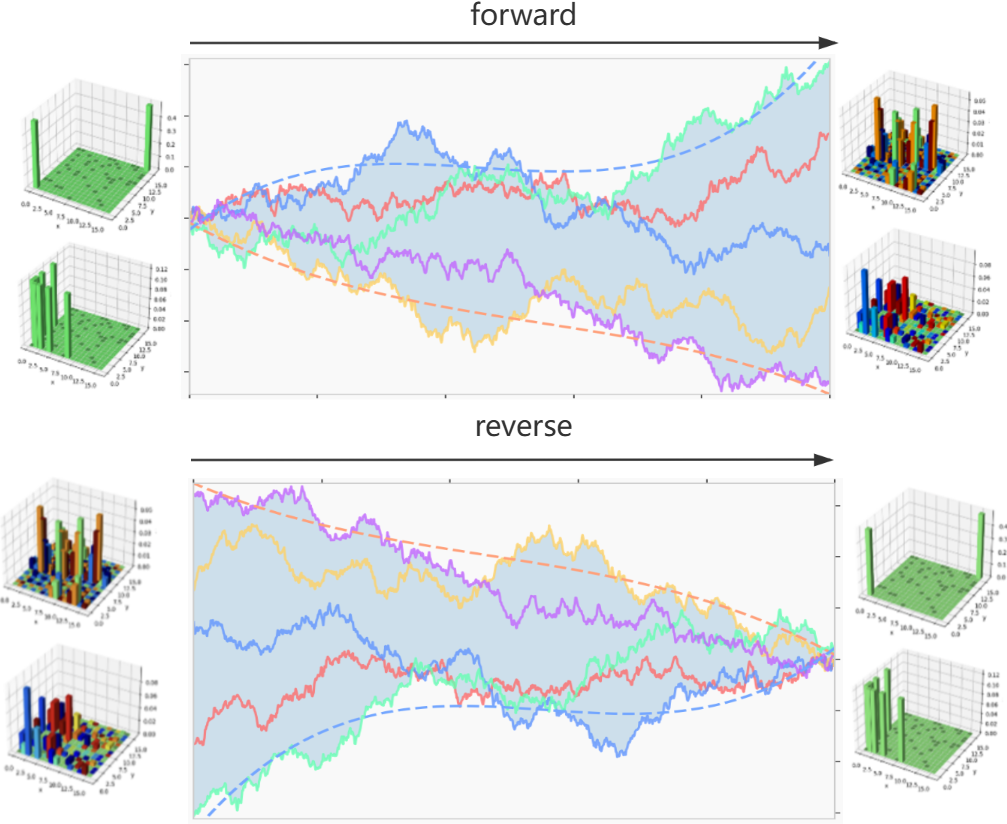}
\caption{Scheme sketch for generating two types of quantum data. } 
\label{fig:sketch-2types}
\end{figure}

\subsection{Denosing the multi types of noisy data simultaneously using our denoising scheme}
Despite the quantum data generation, our denoising models can also be used as a general model to denoise multi-types of quantum data simultaneously. Here, we investigate the performance of our denoise model in scenarios where both GHZ-like and W-like noisy data can be processed using the same network. The sketch of this scenario is shown in Fig.~\ref{fig:sketch-2types}.  Specifically, to engineer the noisy data with different noise levels we artificially add the noise to the GHZ-like and W-like states. We investigate three types of noise environments to test our denoising scheme, the first one $\epsilon_\text{I}$ is the single qubit Rx/Ry noise, and the other two types of noise environments are as follows,
\begin{align}
\epsilon_\text{II}(\rho)=&(1-p)I\nonumber\\+&\frac{p}{3}(R_x(\theta)\rho R_x(\theta)+R_y(\theta)\rho R_y(\theta)+R_z(\theta)\rho R_z(\theta))
\\
\epsilon_\text{III}(\rho)=&R_x(\theta)R_y(\theta)R_z(\theta)\rho R_x(\theta)R_y(\theta)R_z(\theta),
\end{align} 
where the second type of noise environment $\epsilon_\text{II}$, by fixing $\theta$, it returns to the depolarizing environment  $\epsilon(\rho)=(1-p)I+\frac{p}{3}(X\rho X+Y\rho Y+Z\rho Z),$
the single qubit depolarizing noise. For all the three noise environments, the noise level is noted by $\eta\in(0,1]$ and the noise is added by setting the random angle $\theta$ of the related gates sampling from a uniform distribution $[0,\eta\pi]$.

\begin{figure}[ht]
(a)\includegraphics[width=0.9\columnwidth]{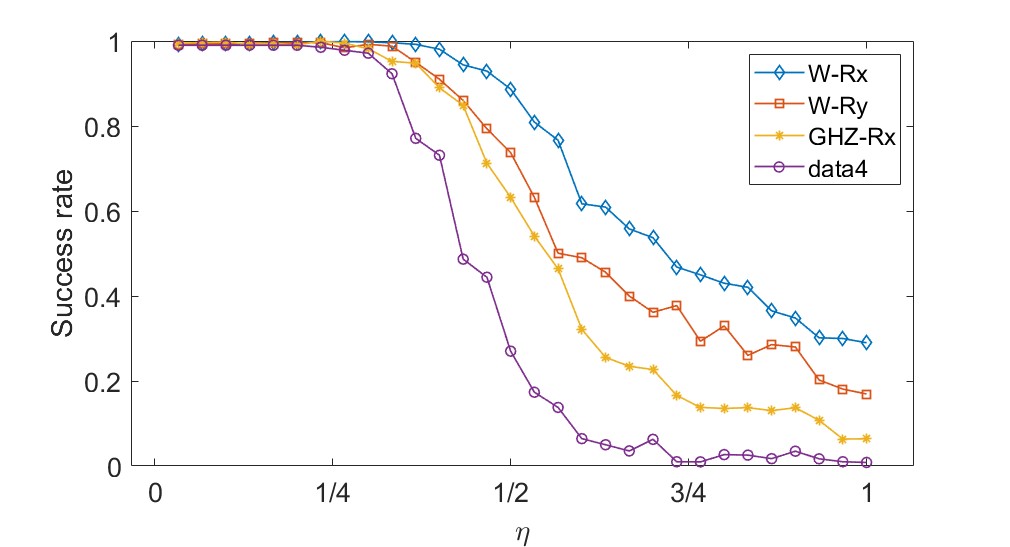}
(b)\includegraphics[width=0.9\columnwidth]{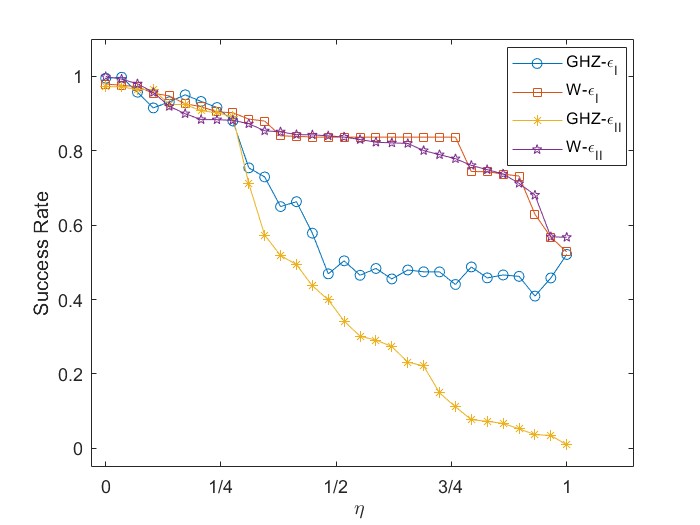}
\caption{The relation between the denoise success rate for two types of data and the single qubit (a) Rx/Ry noise level  $\eta$  and (b) the other two types of noise environments $\epsilon_\text{II}$ and $\epsilon_\text{III}$ in Eqs.(11-12) with $p=0.75$}. 
\label{fig:multi-type-denoising}
\end{figure}

Training our QUNET with two sets of data, i.e. GHZ-like and W-like data, we obtain a denoising model which could serve as a denoising machine for both types of noisy data. The success rate for denoising GHZ-like and W-like data drops gradually with the increase in the noise level $\eta$ and $p=0.75$ as shown in Fig.~\ref{fig:multi-type-denoising}. 

For the first type noise environment $\epsilon_\text{I}$, while the noise level in the single qubit gates is within $[0,1/4]$, the accuracy of the denoised state is close to 100\%. For the noise types $\epsilon_\text{II}$ and $\epsilon_\text{III}$, while the noise level is $\eta<1/4$ and $p=0.75$, the accuracy of the denoised state is above 90\%. The setting of $p=0.75$ is arranged in a large noise level to test the ability of our model since $p\in[0,1]$. These simulations demonstrate the denoising ability of our model.
In our simulations, we use three quantities to evaluate the validity of our method, i.e. signal noise ratio (SNR), mean squared error (MSE), and peak signal noise ratio (PSNR). 
The comparison of the data quality in experiments is shown in Table.~\ref{table:ExpCom}, where we can see the great denoising effect of our scheme. Note that we do not present the case of $\epsilon_\text{II}$ because the model may implement the $I$ operation, which will make the SNR and PSNR meaningless.

\subsection{Implementation detail}
The structure of our QUNET is a bridged MERA structure as shown in Fig.~\ref{qunet-circuit}, which is constructed with two types of operations named disentanglers (the pink square blocks) and isometries (the orange triangle blocks). The disentanglers play the role of convolution layers and the isometries play the role of pooling layers. 
The disentanglers and isometries in our QUNET are the combinations of the single-qubit rotation operators and the specific types of two-qubit operators. The two-qubit operator for disentanglers in convolution layers are the combination of XX,  YY and ZZ gates.
The two-qubit operator for isometries in pooling layers is the CNOT gate. The specific circuit for the disentanglers and isometries in our QUNET are shown in Fig.~\ref{disentangler-isometry}.

\begin{figure}[t]
\includegraphics[width=\columnwidth]{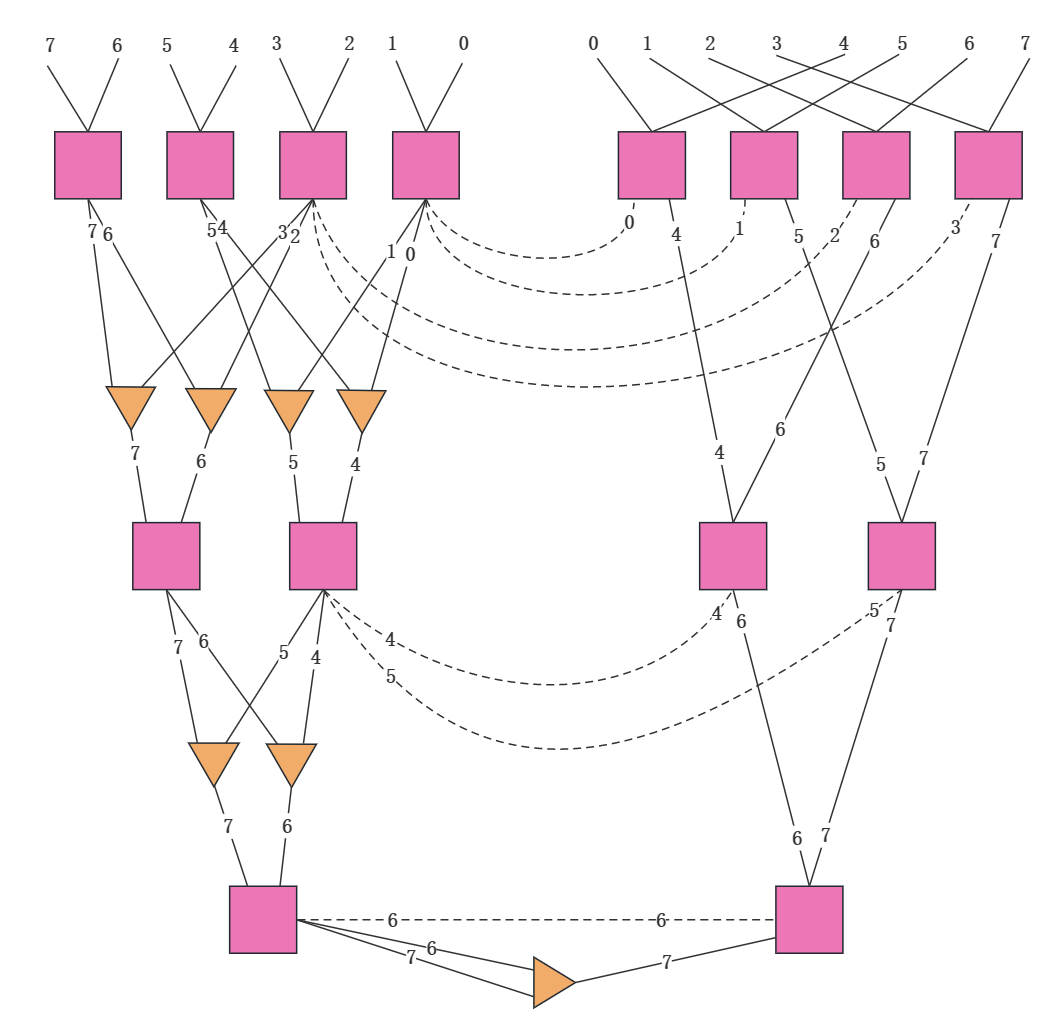}
\caption{The equivalent bridged  MERA representation of the QUNET in Fig.~\ref{qunet-circuit}, with the left part as the downsampling path and the right part as the upsampling path.} 
\label{qunet-mera}
\end{figure}

\begin{figure*}[ht]
\centering
\includegraphics[width=1.9\columnwidth]{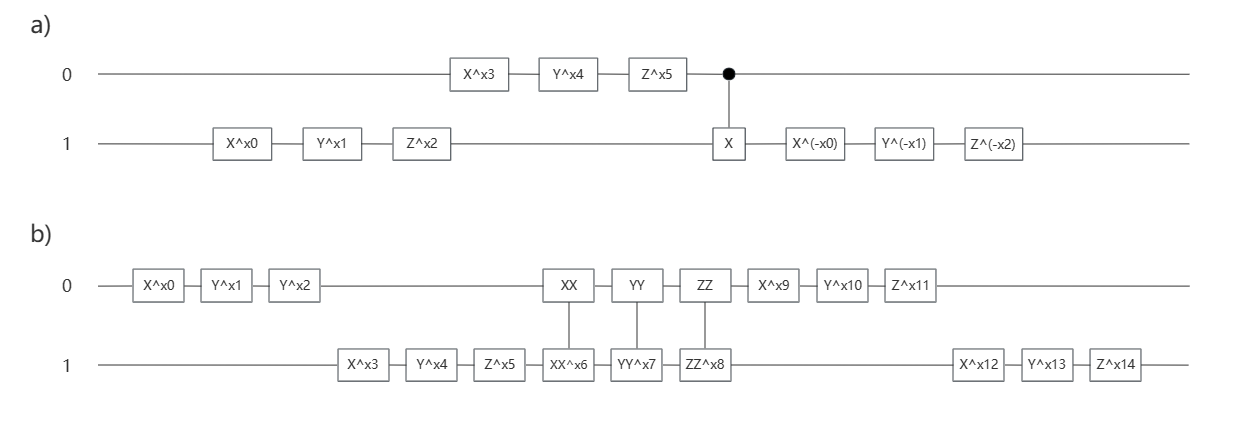}
\caption{The circuit for the realization of (a) isometries and (b) disentanglers in our MERA structure in Fig.~\ref{qunet-mera}.} 
\label{disentangler-isometry}
\end{figure*}

In our QUNET, the number of parameters is linearly dependent on the size of qubits to be generated and the number of quantum convolutional and pooling layers. For the generation of an 8 qubit quantum state, our QUENT shown in Fig.~\ref{qunet-mera} has 112 parameters for the quantum operators therein.

\begin{figure}[t]
\includegraphics[width=\columnwidth]{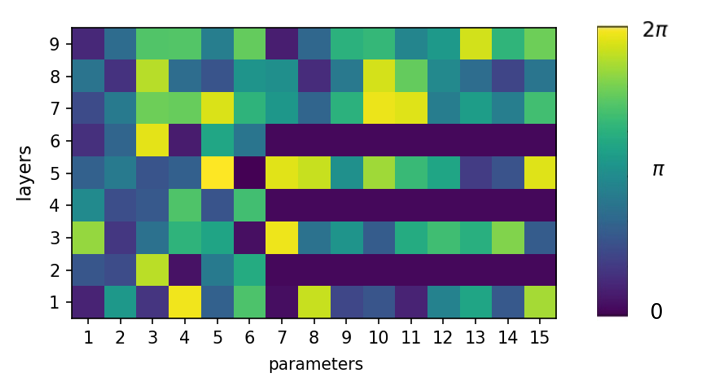}\\
\hspace{-8cm}(a)\\
\includegraphics[width=0.95\columnwidth]{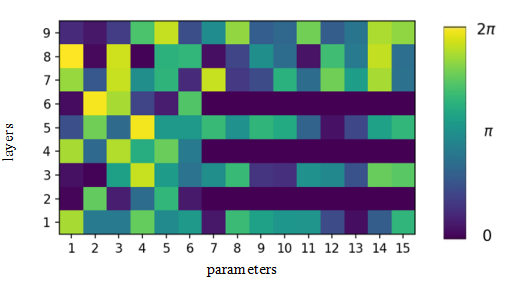}\\
\hspace{-8cm}(b)\\
\caption{The distribution of the parameters in trained QUNET in Fig.~\ref{qunet-circuit} for the generation of GHZ-like states (a), W-like states (b).} 
\label{parameters-trained}
\end{figure}

Considering the trained QUNET for the generation of a specific type of quantum state, the parameters of the corresponding QUENT is the gene of the generation scheme. For the generation of a GHZ-like (W-like) state defined in Eq.\ref{eq:ghz-1} (\ref{eq:w-1}) with QUNET  shown in Fig.~\ref{qunet-circuit}, the distribution of the parameters in trained QUNET is shown in Fig.~\ref{parameters-trained}, where the vertical axis is the layers, and each layer include a different number of quantum operator parameters which is the index of the horizontal axis.

\begin{figure}[h]
\includegraphics[width=0.9\columnwidth]{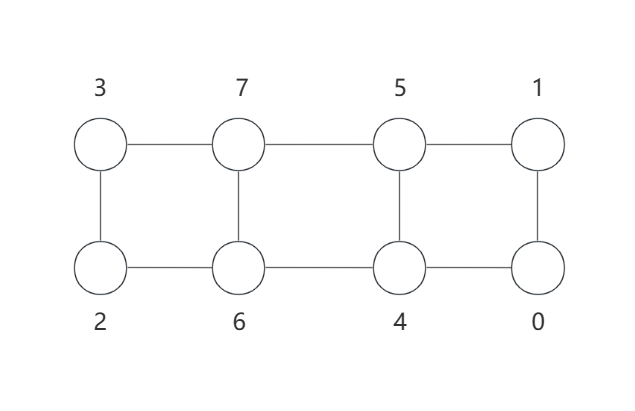}
\caption{The required chip topology for the NISQ quantum processors to realize the QUNET shown in Fig.~\ref{qunet-circuit}.} 
\label{qunet-experiments}
\end{figure}
Our proposal is flexible to design and suitable to be realized in a NISQ processor. With the QUNET for the generation of an 8 qubit state, its corresponding topology of the quantum chip is shown in 
Fig.~\ref{qunet-experiments}, where the available nearest neighbor coupling supports all the two-qubits operators in our QUNET structure in Fig.~\ref{qunet-mera}. This chip topology is a general topology for quantum superconducting chips like USTC's Zuchongzhi quantum processor~\cite{zhu2022quantum} and  Google’s Sycamore quantum processor\cite{arute2019quantum}. Our proposal is designable according to the quantum chip topology and the states to be generated by adjusting the number of layers in our QUNET, the types of quantum operators therein, and the bond dimensions in the equivalent MERA structure.

Our quantum data generation scheme is hardware noise resilience since our scheme is a machine learning-based model and the training process is a dynamic parameters update process. With the existence of all types of hardware noise, the corresponding effect will be the unitary is not exactly the operator we designed.

\begin{figure*}[ht]
\centering
\includegraphics[width=1.9\columnwidth]{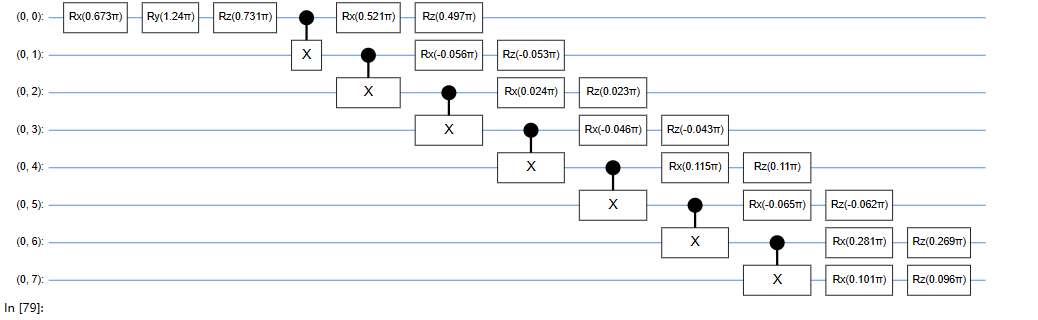}\\\hspace{-16cm}(a)\\
\vspace{1cm}
\includegraphics[width=1.9\columnwidth]{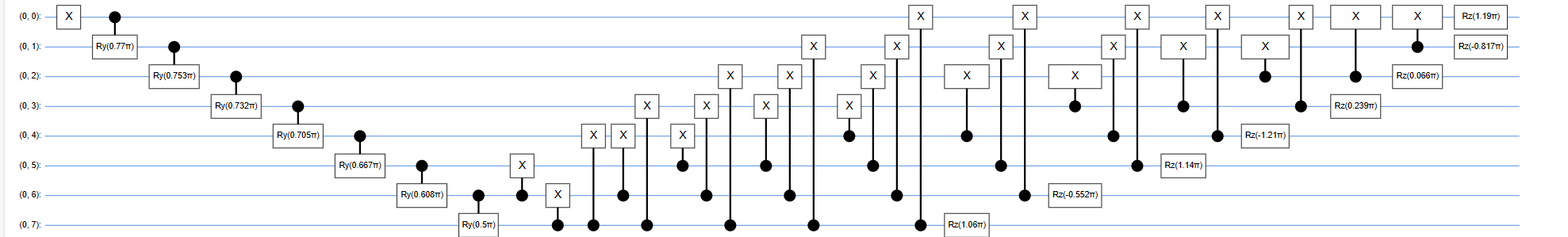}\\\hspace{-16cm}(b)\\
\caption{The circuit for the preparation of training data for (a) GHZ-like state  and (b) W-like state.} 
\label{prepare-circuit}
\end{figure*}

We provide the quantum circuits for the generation of our training data set, as shown in Fig.~\ref{prepare-circuit}. Our simulations are implemented with the TensorFlow quantum package, and the quantum-classical data map is completed with a “Parametrized Quantum Circuit (PQC) Layer”~\cite{benedetti2019parameterized}. 

\section{Conclusion and Discussion}
Our quantum data generation scheme with a denoising process is a flexible efficient method for quantum data preparation, which paves the way for the quantum AI era and sheds light on quantum information processing. Our quantum data generation scheme may also be applied to the study of quantum many-body systems by serving as a system states sampler, and applied for the benchmark of the quantum systems noise. 

Our proposal can be implemented on all types of quantum processors in the NISQ era by shaping the types of disentanglers and isometries in our MERA QUNET. On the one hand, it is easy to switch the disentanglers and isometries operators into the ones easy to realize under the constraint of the quantum chip topology. On the other hand, the bond dimension of MERA structure is also designable, which makes our method flexible and general for NISQ devices. The machine learning intrinsic quality of our scheme enables its noise resistance ability and scalable, which means that the gate fidelity, in reality, will not affect the performance of our model since the training process will learn the gate fidelity-related information also. Our scheme can also be used for denoising of multi-type quantum data by training the network only once. The analytical study of the trained network servers as the map between the full noise state space and the target quantum data space will be an interesting problem to explore.

\section{Data availability}
All data supporting the findings of this study are included in the article (and any supplementary files).

\section{Code availability}
The code is available on request from the authors.
\section{acknowledgements}

 We acknowledge support from the National Natural Science Foundation of China (Grant No.  12104101) and the Fundamental Research Funds for the Central Universities, Stability Program of National Key Laboratory of Security Communication (2023), the Major Research Project of National Natural Science Foundation of China under Grant 92267110, the Joint Funds of the National Natural Science Foundation of China (Grant No. U22B2025) and the Key Research and Development Program of Shaanxi  (Grant No. 2023-GHZD-42).

\section{AUTHOR CONTRIBUTIONS}

X. Huang and L. Shang  performed the experiments. W.-W.Zhang proposed the scheme. W.-W.Zhang, Wei Zhao, B. Yang, W.Pan and H.Shi analyzed the results.  All authors contributed to the writing of the paper.

\section{Competing interests} The authors declare no competing interests.

\bibliographystyle{apsrev4-1}
\bibliography{apssamp}

\begin{thebibliography}{37}%
\makeatletter
\providecommand \@ifxundefined [1]{%
 \@ifx{#1\undefined}
}%
\providecommand \@ifnum [1]{%
 \ifnum #1\expandafter \@firstoftwo
 \else \expandafter \@secondoftwo
 \fi
}%
\providecommand \@ifx [1]{%
 \ifx #1\expandafter \@firstoftwo
 \else \expandafter \@secondoftwo
 \fi
}%
\providecommand \natexlab [1]{#1}%
\providecommand \enquote  [1]{``#1''}%
\providecommand \bibnamefont  [1]{#1}%
\providecommand \bibfnamefont [1]{#1}%
\providecommand \citenamefont [1]{#1}%
\providecommand \href@noop [0]{\@secondoftwo}%
\providecommand \href [0]{\begingroup \@sanitize@url \@href}%
\providecommand \@href[1]{\@@startlink{#1}\@@href}%
\providecommand \@@href[1]{\endgroup#1\@@endlink}%
\providecommand \@sanitize@url [0]{\catcode `\\12\catcode `\$12\catcode `\&12\catcode `\#12\catcode `\^12\catcode `\_12\catcode `\%12\relax}%
\providecommand \@@startlink[1]{}%
\providecommand \@@endlink[0]{}%
\providecommand \url  [0]{\begingroup\@sanitize@url \@url }%
\providecommand \@url [1]{\endgroup\@href {#1}{\urlprefix }}%
\providecommand \urlprefix  [0]{URL }%
\providecommand \Eprint [0]{\href }%
\providecommand \doibase [0]{http://dx.doi.org/}%
\providecommand \selectlanguage [0]{\@gobble}%
\providecommand \bibinfo  [0]{\@secondoftwo}%
\providecommand \bibfield  [0]{\@secondoftwo}%
\providecommand \translation [1]{[#1]}%
\providecommand \BibitemOpen [0]{}%
\providecommand \bibitemStop [0]{}%
\providecommand \bibitemNoStop [0]{.\EOS\space}%
\providecommand \EOS [0]{\spacefactor3000\relax}%
\providecommand \BibitemShut  [1]{\csname bibitem#1\endcsname}%
\let\auto@bib@innerbib\@empty
\bibitem [{\citenamefont {Preskill}(2018)}]{preskill2018quantum}%
  \BibitemOpen
  \bibfield  {author} {\bibinfo {author} {\bibfnamefont {J.}~\bibnamefont {Preskill}},\ }\href@noop {} {\bibfield  {journal} {\bibinfo  {journal} {Quantum}\ }\textbf {\bibinfo {volume} {2}},\ \bibinfo {pages} {79} (\bibinfo {year} {2018})}\BibitemShut {NoStop}%
\bibitem [{\citenamefont {Arute}\ \emph {et~al.}(2019)\citenamefont {Arute}, \citenamefont {Arya}, \citenamefont {Babbush}, \citenamefont {Bacon}, \citenamefont {Bardin}, \citenamefont {Barends}, \citenamefont {Biswas}, \citenamefont {Boixo}, \citenamefont {Brandao}, \citenamefont {Buell} \emph {et~al.}}]{arute2019quantum}%
  \BibitemOpen
  \bibfield  {author} {\bibinfo {author} {\bibfnamefont {F.}~\bibnamefont {Arute}}, \bibinfo {author} {\bibfnamefont {K.}~\bibnamefont {Arya}}, \bibinfo {author} {\bibfnamefont {R.}~\bibnamefont {Babbush}}, \bibinfo {author} {\bibfnamefont {D.}~\bibnamefont {Bacon}}, \bibinfo {author} {\bibfnamefont {J.~C.}\ \bibnamefont {Bardin}}, \bibinfo {author} {\bibfnamefont {R.}~\bibnamefont {Barends}}, \bibinfo {author} {\bibfnamefont {R.}~\bibnamefont {Biswas}}, \bibinfo {author} {\bibfnamefont {S.}~\bibnamefont {Boixo}}, \bibinfo {author} {\bibfnamefont {F.~G.}\ \bibnamefont {Brandao}}, \bibinfo {author} {\bibfnamefont {D.~A.}\ \bibnamefont {Buell}},  \emph {et~al.},\ }\href@noop {} {\bibfield  {journal} {\bibinfo  {journal} {Nature}\ }\textbf {\bibinfo {volume} {574}},\ \bibinfo {pages} {505} (\bibinfo {year} {2019})}\BibitemShut {NoStop}%
\bibitem [{\citenamefont {Zhong}\ \emph {et~al.}(2020)\citenamefont {Zhong}, \citenamefont {Wang}, \citenamefont {Deng}, \citenamefont {Chen}, \citenamefont {Peng}, \citenamefont {Luo}, \citenamefont {Qin}, \citenamefont {Wu}, \citenamefont {Ding}, \citenamefont {Hu} \emph {et~al.}}]{zhong2020quantum}%
  \BibitemOpen
  \bibfield  {author} {\bibinfo {author} {\bibfnamefont {H.-S.}\ \bibnamefont {Zhong}}, \bibinfo {author} {\bibfnamefont {H.}~\bibnamefont {Wang}}, \bibinfo {author} {\bibfnamefont {Y.-H.}\ \bibnamefont {Deng}}, \bibinfo {author} {\bibfnamefont {M.-C.}\ \bibnamefont {Chen}}, \bibinfo {author} {\bibfnamefont {L.-C.}\ \bibnamefont {Peng}}, \bibinfo {author} {\bibfnamefont {Y.-H.}\ \bibnamefont {Luo}}, \bibinfo {author} {\bibfnamefont {J.}~\bibnamefont {Qin}}, \bibinfo {author} {\bibfnamefont {D.}~\bibnamefont {Wu}}, \bibinfo {author} {\bibfnamefont {X.}~\bibnamefont {Ding}}, \bibinfo {author} {\bibfnamefont {Y.}~\bibnamefont {Hu}},  \emph {et~al.},\ }\href@noop {} {\bibfield  {journal} {\bibinfo  {journal} {Science}\ }\textbf {\bibinfo {volume} {370}},\ \bibinfo {pages} {1460} (\bibinfo {year} {2020})}\BibitemShut {NoStop}%
\bibitem [{\citenamefont {Baldwin}\ \emph {et~al.}(2022)\citenamefont {Baldwin}, \citenamefont {Mayer}, \citenamefont {Brown}, \citenamefont {Ryan-Anderson},\ and\ \citenamefont {Hayes}}]{baldwin2022re}%
  \BibitemOpen
  \bibfield  {author} {\bibinfo {author} {\bibfnamefont {C.~H.}\ \bibnamefont {Baldwin}}, \bibinfo {author} {\bibfnamefont {K.}~\bibnamefont {Mayer}}, \bibinfo {author} {\bibfnamefont {N.~C.}\ \bibnamefont {Brown}}, \bibinfo {author} {\bibfnamefont {C.}~\bibnamefont {Ryan-Anderson}}, \ and\ \bibinfo {author} {\bibfnamefont {D.}~\bibnamefont {Hayes}},\ }\href@noop {} {\bibfield  {journal} {\bibinfo  {journal} {Quantum}\ }\textbf {\bibinfo {volume} {6}},\ \bibinfo {pages} {707} (\bibinfo {year} {2022})}\BibitemShut {NoStop}%
\bibitem [{\citenamefont {Zhu}\ \emph {et~al.}(2022)\citenamefont {Zhu}, \citenamefont {Cao}, \citenamefont {Chen}, \citenamefont {Chen}, \citenamefont {Chen}, \citenamefont {Chung}, \citenamefont {Deng}, \citenamefont {Du}, \citenamefont {Fan}, \citenamefont {Gong} \emph {et~al.}}]{zhu2022quantum}%
  \BibitemOpen
  \bibfield  {author} {\bibinfo {author} {\bibfnamefont {Q.}~\bibnamefont {Zhu}}, \bibinfo {author} {\bibfnamefont {S.}~\bibnamefont {Cao}}, \bibinfo {author} {\bibfnamefont {F.}~\bibnamefont {Chen}}, \bibinfo {author} {\bibfnamefont {M.-C.}\ \bibnamefont {Chen}}, \bibinfo {author} {\bibfnamefont {X.}~\bibnamefont {Chen}}, \bibinfo {author} {\bibfnamefont {T.-H.}\ \bibnamefont {Chung}}, \bibinfo {author} {\bibfnamefont {H.}~\bibnamefont {Deng}}, \bibinfo {author} {\bibfnamefont {Y.}~\bibnamefont {Du}}, \bibinfo {author} {\bibfnamefont {D.}~\bibnamefont {Fan}}, \bibinfo {author} {\bibfnamefont {M.}~\bibnamefont {Gong}},  \emph {et~al.},\ }\href@noop {} {\bibfield  {journal} {\bibinfo  {journal} {Science bulletin}\ }\textbf {\bibinfo {volume} {67}},\ \bibinfo {pages} {240} (\bibinfo {year} {2022})}\BibitemShut {NoStop}%
\bibitem [{\citenamefont {Cao}\ \emph {et~al.}(2023{\natexlab{a}})\citenamefont {Cao}, \citenamefont {Wu}, \citenamefont {Chen}, \citenamefont {Gong}, \citenamefont {Wu}, \citenamefont {Ye}, \citenamefont {Zha}, \citenamefont {Qian}, \citenamefont {Ying}, \citenamefont {Guo} \emph {et~al.}}]{cao2023generation}%
  \BibitemOpen
  \bibfield  {author} {\bibinfo {author} {\bibfnamefont {S.}~\bibnamefont {Cao}}, \bibinfo {author} {\bibfnamefont {B.}~\bibnamefont {Wu}}, \bibinfo {author} {\bibfnamefont {F.}~\bibnamefont {Chen}}, \bibinfo {author} {\bibfnamefont {M.}~\bibnamefont {Gong}}, \bibinfo {author} {\bibfnamefont {Y.}~\bibnamefont {Wu}}, \bibinfo {author} {\bibfnamefont {Y.}~\bibnamefont {Ye}}, \bibinfo {author} {\bibfnamefont {C.}~\bibnamefont {Zha}}, \bibinfo {author} {\bibfnamefont {H.}~\bibnamefont {Qian}}, \bibinfo {author} {\bibfnamefont {C.}~\bibnamefont {Ying}}, \bibinfo {author} {\bibfnamefont {S.}~\bibnamefont {Guo}},  \emph {et~al.},\ }\href@noop {} {\bibfield  {journal} {\bibinfo  {journal} {Nature}\ ,\ \bibinfo {pages} {1}} (\bibinfo {year} {2023}{\natexlab{a}})}\BibitemShut {NoStop}%
\bibitem [{\citenamefont {C{\'o}rcoles}\ \emph {et~al.}(2019)\citenamefont {C{\'o}rcoles}, \citenamefont {Kandala}, \citenamefont {Javadi-Abhari}, \citenamefont {McClure}, \citenamefont {Cross}, \citenamefont {Temme}, \citenamefont {Nation}, \citenamefont {Steffen},\ and\ \citenamefont {Gambetta}}]{corcoles2019challenges}%
  \BibitemOpen
  \bibfield  {author} {\bibinfo {author} {\bibfnamefont {A.~D.}\ \bibnamefont {C{\'o}rcoles}}, \bibinfo {author} {\bibfnamefont {A.}~\bibnamefont {Kandala}}, \bibinfo {author} {\bibfnamefont {A.}~\bibnamefont {Javadi-Abhari}}, \bibinfo {author} {\bibfnamefont {D.~T.}\ \bibnamefont {McClure}}, \bibinfo {author} {\bibfnamefont {A.~W.}\ \bibnamefont {Cross}}, \bibinfo {author} {\bibfnamefont {K.}~\bibnamefont {Temme}}, \bibinfo {author} {\bibfnamefont {P.~D.}\ \bibnamefont {Nation}}, \bibinfo {author} {\bibfnamefont {M.}~\bibnamefont {Steffen}}, \ and\ \bibinfo {author} {\bibfnamefont {J.~M.}\ \bibnamefont {Gambetta}},\ }\href@noop {} {\bibfield  {journal} {\bibinfo  {journal} {Proceedings of the IEEE}\ }\textbf {\bibinfo {volume} {108}},\ \bibinfo {pages} {1338} (\bibinfo {year} {2019})}\BibitemShut {NoStop}%
\bibitem [{\citenamefont {Yi}\ \emph {et~al.}(2024)\citenamefont {Yi}, \citenamefont {Hai}, \citenamefont {Luo}, \citenamefont {Zhang}, \citenamefont {Zhou}, \citenamefont {Song}, \citenamefont {Yan}, \citenamefont {Deng},\ and\ \citenamefont {Chen}}]{yi2024robust}%
  \BibitemOpen
  \bibfield  {author} {\bibinfo {author} {\bibfnamefont {K.}~\bibnamefont {Yi}}, \bibinfo {author} {\bibfnamefont {Y.-J.}\ \bibnamefont {Hai}}, \bibinfo {author} {\bibfnamefont {K.}~\bibnamefont {Luo}}, \bibinfo {author} {\bibfnamefont {L.}~\bibnamefont {Zhang}}, \bibinfo {author} {\bibfnamefont {Y.}~\bibnamefont {Zhou}}, \bibinfo {author} {\bibfnamefont {Y.}~\bibnamefont {Song}}, \bibinfo {author} {\bibfnamefont {T.}~\bibnamefont {Yan}}, \bibinfo {author} {\bibfnamefont {X.-H.}\ \bibnamefont {Deng}}, \ and\ \bibinfo {author} {\bibfnamefont {Y.}~\bibnamefont {Chen}},\ }\href@noop {} {\bibfield  {journal} {\bibinfo  {journal} {arXiv preprint arXiv:2401.01810}\ } (\bibinfo {year} {2024})}\BibitemShut {NoStop}%
\bibitem [{\citenamefont {Kolesnikow}\ \emph {et~al.}(2024)\citenamefont {Kolesnikow}, \citenamefont {Bomantara}, \citenamefont {Doherty},\ and\ \citenamefont {Grimsmo}}]{kolesnikow2024gottesman}%
  \BibitemOpen
  \bibfield  {author} {\bibinfo {author} {\bibfnamefont {X.~C.}\ \bibnamefont {Kolesnikow}}, \bibinfo {author} {\bibfnamefont {R.~W.}\ \bibnamefont {Bomantara}}, \bibinfo {author} {\bibfnamefont {A.~C.}\ \bibnamefont {Doherty}}, \ and\ \bibinfo {author} {\bibfnamefont {A.~L.}\ \bibnamefont {Grimsmo}},\ }\href@noop {} {\bibfield  {journal} {\bibinfo  {journal} {Physical Review Letters}\ }\textbf {\bibinfo {volume} {132}},\ \bibinfo {pages} {130605} (\bibinfo {year} {2024})}\BibitemShut {NoStop}%
\bibitem [{\citenamefont {Zhang}\ \emph {et~al.}(2016)\citenamefont {Zhang}, \citenamefont {Goyal}, \citenamefont {Gao}, \citenamefont {Sanders},\ and\ \citenamefont {Simon}}]{catstates-ww}%
  \BibitemOpen
  \bibfield  {author} {\bibinfo {author} {\bibfnamefont {W.-W.}\ \bibnamefont {Zhang}}, \bibinfo {author} {\bibfnamefont {S.~K.}\ \bibnamefont {Goyal}}, \bibinfo {author} {\bibfnamefont {F.}~\bibnamefont {Gao}}, \bibinfo {author} {\bibfnamefont {B.~C.}\ \bibnamefont {Sanders}}, \ and\ \bibinfo {author} {\bibfnamefont {C.}~\bibnamefont {Simon}},\ }\href@noop {} {\bibfield  {journal} {\bibinfo  {journal} {New Journal of Physics}\ }\textbf {\bibinfo {volume} {18}},\ \bibinfo {pages} {093025} (\bibinfo {year} {2016})}\BibitemShut {NoStop}%
\bibitem [{\citenamefont {Yuan}\ and\ \citenamefont {Zhang}(2023{\natexlab{a}})}]{Yuan2023optimalcontrolled}%
  \BibitemOpen
  \bibfield  {author} {\bibinfo {author} {\bibfnamefont {P.}~\bibnamefont {Yuan}}\ and\ \bibinfo {author} {\bibfnamefont {S.}~\bibnamefont {Zhang}},\ }\href {\doibase 10.22331/q-2023-03-20-956} {\bibfield  {journal} {\bibinfo  {journal} {{Quantum}}\ }\textbf {\bibinfo {volume} {7}},\ \bibinfo {pages} {956} (\bibinfo {year} {2023}{\natexlab{a}})}\BibitemShut {NoStop}%
\bibitem [{\citenamefont {Hormozi}\ \emph {et~al.}(2007)\citenamefont {Hormozi}, \citenamefont {Zikos}, \citenamefont {Bonesteel},\ and\ \citenamefont {Simon}}]{hormozi2007topological}%
  \BibitemOpen
  \bibfield  {author} {\bibinfo {author} {\bibfnamefont {L.}~\bibnamefont {Hormozi}}, \bibinfo {author} {\bibfnamefont {G.}~\bibnamefont {Zikos}}, \bibinfo {author} {\bibfnamefont {N.~E.}\ \bibnamefont {Bonesteel}}, \ and\ \bibinfo {author} {\bibfnamefont {S.~H.}\ \bibnamefont {Simon}},\ }\href@noop {} {\bibfield  {journal} {\bibinfo  {journal} {Physical Review B}\ }\textbf {\bibinfo {volume} {75}},\ \bibinfo {pages} {165310} (\bibinfo {year} {2007})}\BibitemShut {NoStop}%
\bibitem [{\citenamefont {Almudever}\ \emph {et~al.}(2017)\citenamefont {Almudever}, \citenamefont {Lao}, \citenamefont {Fu}, \citenamefont {Khammassi}, \citenamefont {Ashraf}, \citenamefont {Iorga}, \citenamefont {Varsamopoulos}, \citenamefont {Eichler}, \citenamefont {Wallraff}, \citenamefont {Geck} \emph {et~al.}}]{almudever2017engineering}%
  \BibitemOpen
  \bibfield  {author} {\bibinfo {author} {\bibfnamefont {C.~G.}\ \bibnamefont {Almudever}}, \bibinfo {author} {\bibfnamefont {L.}~\bibnamefont {Lao}}, \bibinfo {author} {\bibfnamefont {X.}~\bibnamefont {Fu}}, \bibinfo {author} {\bibfnamefont {N.}~\bibnamefont {Khammassi}}, \bibinfo {author} {\bibfnamefont {I.}~\bibnamefont {Ashraf}}, \bibinfo {author} {\bibfnamefont {D.}~\bibnamefont {Iorga}}, \bibinfo {author} {\bibfnamefont {S.}~\bibnamefont {Varsamopoulos}}, \bibinfo {author} {\bibfnamefont {C.}~\bibnamefont {Eichler}}, \bibinfo {author} {\bibfnamefont {A.}~\bibnamefont {Wallraff}}, \bibinfo {author} {\bibfnamefont {L.}~\bibnamefont {Geck}},  \emph {et~al.},\ }in\ \href@noop {} {\emph {\bibinfo {booktitle} {Design, Automation \& Test in Europe Conference \& Exhibition (DATE), 2017}}}\ (\bibinfo {organization} {IEEE},\ \bibinfo {year} {2017})\ pp.\ \bibinfo {pages} {836--845}\BibitemShut {NoStop}%
\bibitem [{\citenamefont {Venturelli}\ \emph {et~al.}(2019)\citenamefont {Venturelli}, \citenamefont {Do}, \citenamefont {O'Gorman}, \citenamefont {Frank}, \citenamefont {Rieffel}, \citenamefont {Booth}, \citenamefont {Nguyen}, \citenamefont {Narayan},\ and\ \citenamefont {Nanda}}]{venturelli2019quantum}%
  \BibitemOpen
  \bibfield  {author} {\bibinfo {author} {\bibfnamefont {D.}~\bibnamefont {Venturelli}}, \bibinfo {author} {\bibfnamefont {M.}~\bibnamefont {Do}}, \bibinfo {author} {\bibfnamefont {B.}~\bibnamefont {O'Gorman}}, \bibinfo {author} {\bibfnamefont {J.}~\bibnamefont {Frank}}, \bibinfo {author} {\bibfnamefont {E.}~\bibnamefont {Rieffel}}, \bibinfo {author} {\bibfnamefont {K.~E.}\ \bibnamefont {Booth}}, \bibinfo {author} {\bibfnamefont {T.}~\bibnamefont {Nguyen}}, \bibinfo {author} {\bibfnamefont {P.}~\bibnamefont {Narayan}}, \ and\ \bibinfo {author} {\bibfnamefont {S.}~\bibnamefont {Nanda}},\ }in\ \href@noop {} {\emph {\bibinfo {booktitle} {Scheduling and Planning Applications Workshop}}}\ (\bibinfo {year} {2019})\BibitemShut {NoStop}%
\bibitem [{\citenamefont {Sharma}\ \emph {et~al.}(2020)\citenamefont {Sharma}, \citenamefont {Khatri}, \citenamefont {Cerezo},\ and\ \citenamefont {Coles}}]{sharma2020noise}%
  \BibitemOpen
  \bibfield  {author} {\bibinfo {author} {\bibfnamefont {K.}~\bibnamefont {Sharma}}, \bibinfo {author} {\bibfnamefont {S.}~\bibnamefont {Khatri}}, \bibinfo {author} {\bibfnamefont {M.}~\bibnamefont {Cerezo}}, \ and\ \bibinfo {author} {\bibfnamefont {P.~J.}\ \bibnamefont {Coles}},\ }\href@noop {} {\bibfield  {journal} {\bibinfo  {journal} {New Journal of Physics}\ }\textbf {\bibinfo {volume} {22}},\ \bibinfo {pages} {043006} (\bibinfo {year} {2020})}\BibitemShut {NoStop}%
\bibitem [{\citenamefont {Gleinig}\ and\ \citenamefont {Hoefler}(2021)}]{gleinig2021efficient}%
  \BibitemOpen
  \bibfield  {author} {\bibinfo {author} {\bibfnamefont {N.}~\bibnamefont {Gleinig}}\ and\ \bibinfo {author} {\bibfnamefont {T.}~\bibnamefont {Hoefler}},\ }in\ \href@noop {} {\emph {\bibinfo {booktitle} {2021 58th ACM/IEEE Design Automation Conference (DAC)}}}\ (\bibinfo {organization} {IEEE},\ \bibinfo {year} {2021})\ pp.\ \bibinfo {pages} {433--438}\BibitemShut {NoStop}%
\bibitem [{\citenamefont {Zhang}\ \emph {et~al.}(2022)\citenamefont {Zhang}, \citenamefont {Li},\ and\ \citenamefont {Yuan}}]{zhang2022quantum}%
  \BibitemOpen
  \bibfield  {author} {\bibinfo {author} {\bibfnamefont {X.-M.}\ \bibnamefont {Zhang}}, \bibinfo {author} {\bibfnamefont {T.}~\bibnamefont {Li}}, \ and\ \bibinfo {author} {\bibfnamefont {X.}~\bibnamefont {Yuan}},\ }\href@noop {} {\bibfield  {journal} {\bibinfo  {journal} {Physical Review Letters}\ }\textbf {\bibinfo {volume} {129}},\ \bibinfo {pages} {230504} (\bibinfo {year} {2022})}\BibitemShut {NoStop}%
\bibitem [{\citenamefont {Yuan}\ and\ \citenamefont {Zhang}(2023{\natexlab{b}})}]{yuan2023optimal}%
  \BibitemOpen
  \bibfield  {author} {\bibinfo {author} {\bibfnamefont {P.}~\bibnamefont {Yuan}}\ and\ \bibinfo {author} {\bibfnamefont {S.}~\bibnamefont {Zhang}},\ }\href@noop {} {\bibfield  {journal} {\bibinfo  {journal} {Quantum}\ }\textbf {\bibinfo {volume} {7}},\ \bibinfo {pages} {956} (\bibinfo {year} {2023}{\natexlab{b}})}\BibitemShut {NoStop}%
\bibitem [{\citenamefont {Melnikov}\ \emph {et~al.}(2023)\citenamefont {Melnikov}, \citenamefont {Termanova}, \citenamefont {Dolgov}, \citenamefont {Neukart},\ and\ \citenamefont {Perelshtein}}]{melnikov2023quantum}%
  \BibitemOpen
  \bibfield  {author} {\bibinfo {author} {\bibfnamefont {A.~A.}\ \bibnamefont {Melnikov}}, \bibinfo {author} {\bibfnamefont {A.~A.}\ \bibnamefont {Termanova}}, \bibinfo {author} {\bibfnamefont {S.~V.}\ \bibnamefont {Dolgov}}, \bibinfo {author} {\bibfnamefont {F.}~\bibnamefont {Neukart}}, \ and\ \bibinfo {author} {\bibfnamefont {M.}~\bibnamefont {Perelshtein}},\ }\href@noop {} {\bibfield  {journal} {\bibinfo  {journal} {Quantum Science and Technology}\ } (\bibinfo {year} {2023})}\BibitemShut {NoStop}%
\bibitem [{\citenamefont {Lu}\ \emph {et~al.}(2023)\citenamefont {Lu}, \citenamefont {Zhou}, \citenamefont {Fei},\ and\ \citenamefont {Ran}}]{lu2023quantum}%
  \BibitemOpen
  \bibfield  {author} {\bibinfo {author} {\bibfnamefont {Y.}~\bibnamefont {Lu}}, \bibinfo {author} {\bibfnamefont {P.-F.}\ \bibnamefont {Zhou}}, \bibinfo {author} {\bibfnamefont {S.-M.}\ \bibnamefont {Fei}}, \ and\ \bibinfo {author} {\bibfnamefont {S.-J.}\ \bibnamefont {Ran}},\ }\href@noop {} {\bibfield  {journal} {\bibinfo  {journal} {Physical Review Research}\ }\textbf {\bibinfo {volume} {5}},\ \bibinfo {pages} {023096} (\bibinfo {year} {2023})}\BibitemShut {NoStop}%
\bibitem [{\citenamefont {Porotti}\ \emph {et~al.}(2022)\citenamefont {Porotti}, \citenamefont {Essig}, \citenamefont {Huard},\ and\ \citenamefont {Marquardt}}]{porotti2022deep}%
  \BibitemOpen
  \bibfield  {author} {\bibinfo {author} {\bibfnamefont {R.}~\bibnamefont {Porotti}}, \bibinfo {author} {\bibfnamefont {A.}~\bibnamefont {Essig}}, \bibinfo {author} {\bibfnamefont {B.}~\bibnamefont {Huard}}, \ and\ \bibinfo {author} {\bibfnamefont {F.}~\bibnamefont {Marquardt}},\ }\href@noop {} {\bibfield  {journal} {\bibinfo  {journal} {Quantum}\ }\textbf {\bibinfo {volume} {6}},\ \bibinfo {pages} {747} (\bibinfo {year} {2022})}\BibitemShut {NoStop}%
\bibitem [{\citenamefont {Li}\ \emph {et~al.}(2023)\citenamefont {Li}, \citenamefont {He},\ and\ \citenamefont {Wang}}]{PhysRevA.108.052418}%
  \BibitemOpen
  \bibfield  {author} {\bibinfo {author} {\bibfnamefont {C.-C.}\ \bibnamefont {Li}}, \bibinfo {author} {\bibfnamefont {R.-H.}\ \bibnamefont {He}}, \ and\ \bibinfo {author} {\bibfnamefont {Z.-M.}\ \bibnamefont {Wang}},\ }\href {\doibase 10.1103/PhysRevA.108.052418} {\bibfield  {journal} {\bibinfo  {journal} {Phys. Rev. A}\ }\textbf {\bibinfo {volume} {108}},\ \bibinfo {pages} {052418} (\bibinfo {year} {2023})}\BibitemShut {NoStop}%
\bibitem [{\citenamefont {Mackeprang}\ \emph {et~al.}(2020)\citenamefont {Mackeprang}, \citenamefont {Dasari},\ and\ \citenamefont {Wrachtrup}}]{mackeprang2020reinforcement}%
  \BibitemOpen
  \bibfield  {author} {\bibinfo {author} {\bibfnamefont {J.}~\bibnamefont {Mackeprang}}, \bibinfo {author} {\bibfnamefont {D.~B.~R.}\ \bibnamefont {Dasari}}, \ and\ \bibinfo {author} {\bibfnamefont {J.}~\bibnamefont {Wrachtrup}},\ }\href@noop {} {\bibfield  {journal} {\bibinfo  {journal} {Quantum Machine Intelligence}\ }\textbf {\bibinfo {volume} {2}},\ \bibinfo {pages} {1} (\bibinfo {year} {2020})}\BibitemShut {NoStop}%
\bibitem [{\citenamefont {Zhang}\ \emph {et~al.}(2019)\citenamefont {Zhang}, \citenamefont {Wei}, \citenamefont {Asad}, \citenamefont {Yang},\ and\ \citenamefont {Wang}}]{zhang2019does}%
  \BibitemOpen
  \bibfield  {author} {\bibinfo {author} {\bibfnamefont {X.-M.}\ \bibnamefont {Zhang}}, \bibinfo {author} {\bibfnamefont {Z.}~\bibnamefont {Wei}}, \bibinfo {author} {\bibfnamefont {R.}~\bibnamefont {Asad}}, \bibinfo {author} {\bibfnamefont {X.-C.}\ \bibnamefont {Yang}}, \ and\ \bibinfo {author} {\bibfnamefont {X.}~\bibnamefont {Wang}},\ }\href@noop {} {\bibfield  {journal} {\bibinfo  {journal} {npj Quantum Information}\ }\textbf {\bibinfo {volume} {5}},\ \bibinfo {pages} {85} (\bibinfo {year} {2019})}\BibitemShut {NoStop}%
\bibitem [{\citenamefont {Arrazola}\ \emph {et~al.}(2019)\citenamefont {Arrazola}, \citenamefont {Bromley}, \citenamefont {Izaac}, \citenamefont {Myers}, \citenamefont {Br{\'a}dler},\ and\ \citenamefont {Killoran}}]{arrazola2019machine}%
  \BibitemOpen
  \bibfield  {author} {\bibinfo {author} {\bibfnamefont {J.~M.}\ \bibnamefont {Arrazola}}, \bibinfo {author} {\bibfnamefont {T.~R.}\ \bibnamefont {Bromley}}, \bibinfo {author} {\bibfnamefont {J.}~\bibnamefont {Izaac}}, \bibinfo {author} {\bibfnamefont {C.~R.}\ \bibnamefont {Myers}}, \bibinfo {author} {\bibfnamefont {K.}~\bibnamefont {Br{\'a}dler}}, \ and\ \bibinfo {author} {\bibfnamefont {N.}~\bibnamefont {Killoran}},\ }\href@noop {} {\bibfield  {journal} {\bibinfo  {journal} {Quantum Science and Technology}\ }\textbf {\bibinfo {volume} {4}},\ \bibinfo {pages} {024004} (\bibinfo {year} {2019})}\BibitemShut {NoStop}%
\bibitem [{\citenamefont {Bahrini}\ \emph {et~al.}(2023)\citenamefont {Bahrini}, \citenamefont {Khamoshifar}, \citenamefont {Abbasimehr}, \citenamefont {Riggs}, \citenamefont {Esmaeili}, \citenamefont {Majdabadkohne},\ and\ \citenamefont {Pasehvar}}]{10137850}%
  \BibitemOpen
  \bibfield  {author} {\bibinfo {author} {\bibfnamefont {A.}~\bibnamefont {Bahrini}}, \bibinfo {author} {\bibfnamefont {M.}~\bibnamefont {Khamoshifar}}, \bibinfo {author} {\bibfnamefont {H.}~\bibnamefont {Abbasimehr}}, \bibinfo {author} {\bibfnamefont {R.~J.}\ \bibnamefont {Riggs}}, \bibinfo {author} {\bibfnamefont {M.}~\bibnamefont {Esmaeili}}, \bibinfo {author} {\bibfnamefont {R.~M.}\ \bibnamefont {Majdabadkohne}}, \ and\ \bibinfo {author} {\bibfnamefont {M.}~\bibnamefont {Pasehvar}},\ }in\ \href {\doibase 10.1109/SIEDS58326.2023.10137850} {\emph {\bibinfo {booktitle} {2023 Systems and Information Engineering Design Symposium (SIEDS)}}}\ (\bibinfo {year} {2023})\ pp.\ \bibinfo {pages} {274--279}\BibitemShut {NoStop}%
\bibitem [{\citenamefont {Cao}\ \emph {et~al.}(2023{\natexlab{b}})\citenamefont {Cao}, \citenamefont {Tan}, \citenamefont {Gao}, \citenamefont {Xu}, \citenamefont {Chen}, \citenamefont {Heng},\ and\ \citenamefont {Li}}]{cao2023survey}%
  \BibitemOpen
  \bibfield  {author} {\bibinfo {author} {\bibfnamefont {H.}~\bibnamefont {Cao}}, \bibinfo {author} {\bibfnamefont {C.}~\bibnamefont {Tan}}, \bibinfo {author} {\bibfnamefont {Z.}~\bibnamefont {Gao}}, \bibinfo {author} {\bibfnamefont {Y.}~\bibnamefont {Xu}}, \bibinfo {author} {\bibfnamefont {G.}~\bibnamefont {Chen}}, \bibinfo {author} {\bibfnamefont {P.-A.}\ \bibnamefont {Heng}}, \ and\ \bibinfo {author} {\bibfnamefont {S.~Z.}\ \bibnamefont {Li}},\ }\href@noop {} {\enquote {\bibinfo {title} {A survey on generative diffusion model},}\ } (\bibinfo {year} {2023}{\natexlab{b}}),\ \Eprint {http://arxiv.org/abs/2209.02646} {arXiv:2209.02646 [cs.AI]} \BibitemShut {NoStop}%
\bibitem [{\citenamefont {Rezende}\ \emph {et~al.}(2014)\citenamefont {Rezende}, \citenamefont {Mohamed},\ and\ \citenamefont {Wierstra}}]{rezende2014stochastic}%
  \BibitemOpen
  \bibfield  {author} {\bibinfo {author} {\bibfnamefont {D.~J.}\ \bibnamefont {Rezende}}, \bibinfo {author} {\bibfnamefont {S.}~\bibnamefont {Mohamed}}, \ and\ \bibinfo {author} {\bibfnamefont {D.}~\bibnamefont {Wierstra}},\ }in\ \href@noop {} {\emph {\bibinfo {booktitle} {International conference on machine learning}}}\ (\bibinfo {organization} {PMLR},\ \bibinfo {year} {2014})\ pp.\ \bibinfo {pages} {1278--1286}\BibitemShut {NoStop}%
\bibitem [{\citenamefont {Creswell}\ \emph {et~al.}(2018)\citenamefont {Creswell}, \citenamefont {White}, \citenamefont {Dumoulin}, \citenamefont {Arulkumaran}, \citenamefont {Sengupta},\ and\ \citenamefont {Bharath}}]{creswell2018generative}%
  \BibitemOpen
  \bibfield  {author} {\bibinfo {author} {\bibfnamefont {A.}~\bibnamefont {Creswell}}, \bibinfo {author} {\bibfnamefont {T.}~\bibnamefont {White}}, \bibinfo {author} {\bibfnamefont {V.}~\bibnamefont {Dumoulin}}, \bibinfo {author} {\bibfnamefont {K.}~\bibnamefont {Arulkumaran}}, \bibinfo {author} {\bibfnamefont {B.}~\bibnamefont {Sengupta}}, \ and\ \bibinfo {author} {\bibfnamefont {A.~A.}\ \bibnamefont {Bharath}},\ }\href@noop {} {\bibfield  {journal} {\bibinfo  {journal} {IEEE signal processing magazine}\ }\textbf {\bibinfo {volume} {35}},\ \bibinfo {pages} {53} (\bibinfo {year} {2018})}\BibitemShut {NoStop}%
\bibitem [{\citenamefont {Rezende}\ and\ \citenamefont {Mohamed}(2015)}]{rezende2015variational}%
  \BibitemOpen
  \bibfield  {author} {\bibinfo {author} {\bibfnamefont {D.}~\bibnamefont {Rezende}}\ and\ \bibinfo {author} {\bibfnamefont {S.}~\bibnamefont {Mohamed}},\ }in\ \href@noop {} {\emph {\bibinfo {booktitle} {International conference on machine learning}}}\ (\bibinfo {organization} {PMLR},\ \bibinfo {year} {2015})\ pp.\ \bibinfo {pages} {1530--1538}\BibitemShut {NoStop}%
\bibitem [{\citenamefont {Ho}\ \emph {et~al.}(2020)\citenamefont {Ho}, \citenamefont {Jain},\ and\ \citenamefont {Abbeel}}]{ho2020denoising}%
  \BibitemOpen
  \bibfield  {author} {\bibinfo {author} {\bibfnamefont {J.}~\bibnamefont {Ho}}, \bibinfo {author} {\bibfnamefont {A.}~\bibnamefont {Jain}}, \ and\ \bibinfo {author} {\bibfnamefont {P.}~\bibnamefont {Abbeel}},\ }\href@noop {} {\bibfield  {journal} {\bibinfo  {journal} {Advances in neural information processing systems}\ }\textbf {\bibinfo {volume} {33}},\ \bibinfo {pages} {6840} (\bibinfo {year} {2020})}\BibitemShut {NoStop}%
\bibitem [{\citenamefont {Nahum}\ \emph {et~al.}(2017)\citenamefont {Nahum}, \citenamefont {Ruhman}, \citenamefont {Vijay},\ and\ \citenamefont {Haah}}]{nahum2017quantum}%
  \BibitemOpen
  \bibfield  {author} {\bibinfo {author} {\bibfnamefont {A.}~\bibnamefont {Nahum}}, \bibinfo {author} {\bibfnamefont {J.}~\bibnamefont {Ruhman}}, \bibinfo {author} {\bibfnamefont {S.}~\bibnamefont {Vijay}}, \ and\ \bibinfo {author} {\bibfnamefont {J.}~\bibnamefont {Haah}},\ }\href@noop {} {\bibfield  {journal} {\bibinfo  {journal} {Physical Review X}\ }\textbf {\bibinfo {volume} {7}},\ \bibinfo {pages} {031016} (\bibinfo {year} {2017})}\BibitemShut {NoStop}%
\bibitem [{\citenamefont {Zhou}\ and\ \citenamefont {Nahum}(2020)}]{zhou2020entanglement}%
  \BibitemOpen
  \bibfield  {author} {\bibinfo {author} {\bibfnamefont {T.}~\bibnamefont {Zhou}}\ and\ \bibinfo {author} {\bibfnamefont {A.}~\bibnamefont {Nahum}},\ }\href@noop {} {\bibfield  {journal} {\bibinfo  {journal} {Physical Review X}\ }\textbf {\bibinfo {volume} {10}},\ \bibinfo {pages} {031066} (\bibinfo {year} {2020})}\BibitemShut {NoStop}%
\bibitem [{\citenamefont {Ronneberger}\ \emph {et~al.}(2015)\citenamefont {Ronneberger}, \citenamefont {Fischer},\ and\ \citenamefont {Brox}}]{ronneberger2015u}%
  \BibitemOpen
  \bibfield  {author} {\bibinfo {author} {\bibfnamefont {O.}~\bibnamefont {Ronneberger}}, \bibinfo {author} {\bibfnamefont {P.}~\bibnamefont {Fischer}}, \ and\ \bibinfo {author} {\bibfnamefont {T.}~\bibnamefont {Brox}},\ }in\ \href@noop {} {\emph {\bibinfo {booktitle} {Medical Image Computing and Computer-Assisted Intervention--MICCAI 2015: 18th International Conference, Munich, Germany, October 5-9, 2015, Proceedings, Part III 18}}}\ (\bibinfo {organization} {Springer},\ \bibinfo {year} {2015})\ pp.\ \bibinfo {pages} {234--241}\BibitemShut {NoStop}%
\bibitem [{\citenamefont {Cong}\ \emph {et~al.}(2019)\citenamefont {Cong}, \citenamefont {Choi},\ and\ \citenamefont {Lukin}}]{cong2019quantum}%
  \BibitemOpen
  \bibfield  {author} {\bibinfo {author} {\bibfnamefont {I.}~\bibnamefont {Cong}}, \bibinfo {author} {\bibfnamefont {S.}~\bibnamefont {Choi}}, \ and\ \bibinfo {author} {\bibfnamefont {M.~D.}\ \bibnamefont {Lukin}},\ }\href@noop {} {\bibfield  {journal} {\bibinfo  {journal} {Nature Physics}\ }\textbf {\bibinfo {volume} {15}},\ \bibinfo {pages} {1273} (\bibinfo {year} {2019})}\BibitemShut {NoStop}%
\bibitem [{\citenamefont {Vidal}(2008)}]{vidal2008class}%
  \BibitemOpen
  \bibfield  {author} {\bibinfo {author} {\bibfnamefont {G.}~\bibnamefont {Vidal}},\ }\href@noop {} {\bibfield  {journal} {\bibinfo  {journal} {Physical review letters}\ }\textbf {\bibinfo {volume} {101}},\ \bibinfo {pages} {110501} (\bibinfo {year} {2008})}\BibitemShut {NoStop}%
\bibitem [{\citenamefont {Benedetti}\ \emph {et~al.}(2019)\citenamefont {Benedetti}, \citenamefont {Lloyd}, \citenamefont {Sack},\ and\ \citenamefont {Fiorentini}}]{benedetti2019parameterized}%
  \BibitemOpen
  \bibfield  {author} {\bibinfo {author} {\bibfnamefont {M.}~\bibnamefont {Benedetti}}, \bibinfo {author} {\bibfnamefont {E.}~\bibnamefont {Lloyd}}, \bibinfo {author} {\bibfnamefont {S.}~\bibnamefont {Sack}}, \ and\ \bibinfo {author} {\bibfnamefont {M.}~\bibnamefont {Fiorentini}},\ }\href@noop {} {\bibfield  {journal} {\bibinfo  {journal} {Quantum Science and Technology}\ }\textbf {\bibinfo {volume} {4}},\ \bibinfo {pages} {043001} (\bibinfo {year} {2019})}\BibitemShut {NoStop}%
\end{thebibliography}%

\end{document}